\documentclass{WileyMSP-template}

\usepackage{amsmath,amssymb}
\usepackage{capt-of}
\usepackage{siunitx}
\usepackage{mathrsfs}
\begin{document}

\title{Low Phase Noise, Record-High Power Optical Parametric Oscillator Tunable from \SIrange[]{2.7}{4.7}{\micro\meter} for Metrology Applications}

\maketitle

% Author: Please give full first and last names for authors and include * after the name of all corresponding authors
\author{Vito F. Pecile\,$^{1,2,*}$,}
\author{Michael Leskowschek\,$^{1}$,}
\author{Norbert Modsching\,$^{3}$,}
\author{Valentin J. Wittwer\,$^{3}$,}
\newline
\author{Thomas Südmeyer\,$^{3}$,}
\author{Oliver H. Heckl\,$^{1,*}$}

% Dedication
%\dedication{Optional dedication here. If no dedication is required, please leave blank}

% Affiliations: Please provide adacemic titles (Prof. or Dr.) for all authors where applicable, and include an institutional email address for all corresponding authors
\begin{affiliations}
$^{1}$\,University of Vienna, Faculty of Physics, Faculty Center for Nano Structure Research, Christian Doppler Laboratory for Mid-IR Spectroscopy, Boltzmanngasse 5, 1090 Vienna, Austria\\
$^{2}$\,University of Vienna, Vienna Doctoral School in Physics, Boltzmanngasse 5, 1090 Vienna, Austria\\
$^{3}$\,Laboratoire Temps-Fréquence (LTF), Institut de Physique, Université de Neuchâtel, Avenue de Bellevaux 51, 2000 Neuchâtel, Switzerland\\
$^{*}$\,vito.pecile@univie.ac.at, oliver.heckl@univie.ac.at\\
\end{affiliations}

% Abstract should be written in the present tense and impersonal style (i.e., avoid we), and be at most 200 words long
\begin{abstract}
\textbf{Abstract:}\\
Within the domain of optical frequency comb systems operating in the mid-infrared, extensive exploration has been undertaken regarding critical parameters such as stabilization, coherence, or spectral tunability. Despite this, certain essential parameters remain inadequately addressed, particularly concerning the light source prerequisites for advanced spectroscopy techniques operating in the \SIrange{3}{5}{\micro\meter} spectral region. Specifically, the necessity for high simultaneous stability of average power, spectral shape, and temporal properties, alongside requirements for excellent beam quality and high powers of several Watts, emerges for applications like cavity-enhanced Doppler-free saturation spectroscopy. Existing systems fall short of meeting these rigorous criteria. This study delves into the metrology aspects of an optical parametric oscillator system, with a particular emphasis on its suitability for power-hungry metrology applications. Notably, the highest average power reported in the \SIrange{3}{5}{\micro\meter} region, reaching \SI{10.3}{W} for the idler output at \SI{3.1}{\micro\meter}, is achieved. Additionally, new aspects, like the analysis of idler phase noise and beam quality and the onset of higher order modes are discussed. These findings represent a significant advancement towards the realization of highly-stable, metrology-grade frequency combs in the mid-infrared, thereby facilitating precision spectroscopy techniques previously constrained by light source average powers and quality limitations.

\end{abstract}

% Text: Please use section headings and subheadings as specified below. For communications, all section headings apart from Experimental Section should be removed
% Please make the first reference to a display item bold: \textbf{Figure 1}
% Do not abbreviate Figure, Equation, etc.; display items are always singular, i.e., Figure 1 and 2.
% Equations are always singular, i.e., Equation 1 and 2, and should be inserted using the {equation} environment, not as graphics
% Please do not use footnotes in the text, additional information can be added to the Reference list.

\section{Introduction}

Recent applications like human breath analysis\,\cite{Metsala2018}, spectroscopy in the "fingerprint region"\,\cite{Changala2019}, or trace-gas sensing\,\cite{Foltynowicz2013} have fostered the development of light sources that expand the optical frequency comb (OFC) technology to the mid-infrared (mid-IR). Several different approaches, among them mode-locked quantum cascade lasers (QCLs)\,\cite{Hugi2012}, or systems based on parametric amplification, like difference frequency generation (DFG)\,\cite{Catanese2019,Maser2017,Hu2017,Murray2016}, optical parametric amplifiers (OPAs)\,\cite{Kanai2019,Manzoni2016,Cerullo2003}, or optical parametric oscillators (OPOs)\,\cite{Bauer2022,Nagele2020,Jornod2017,Coluccelli2017,Balskus2016,Balskus:2016,Lee2014,Ferreiro2011,Adler2009,Sudmeyer2004} have been explored and experienced a huge gain in attention over the past few years. 

While many interesting properties of those systems, among them OFC stabilization\,\cite{Balskus2016,Lee2014,Ferreiro2011,Adler2009}, coherence\,\cite{Maser2017}, spectral tunability\,\cite{Nagele2020,Adler2009}, spectral stability and stabilization\,\cite{Nagele2020,Vainio2017} or output power stability\,\cite{Catanese2019,Nagele2020,Vainio2017}, have already been studied, several vital parameters have not been discussed in sufficient detail or are still completely missing in literature. This becomes particularly evident when looking at the light source requirements of advanced spectroscopy techniques like cavity-enhanced Doppler-free saturation spectroscopy in the of \SIrange[]{3}{5}{\micro\meter} region: Here, frequency metrology aspects, like a high simultaneous stability of average power, the spectral shape, as well as temporal properties (i.e., the phase noise (PN)) are important, while at the same time excellent beam quality (i.e., low M$^2$) at high average powers of up to several Watts is demanded\,\cite{Shumakova2024}. Additionally, the idler spectra should be tunable over a wide range to access different molecular absorption features, and have a narrow bandwidth (BW) on the order of \SIrange{50}{100}{nm} full width at half maximum (FWHM), to match the reflective band of state-of-the-art low-loss mirrors in the mid-IR\,\cite{Truong2023}, which are used in the enhancement cavity.

Comparing these demands with state-of-the-art systems in the \SIrange{3}{5}{\micro\meter} region and their discussed properties, we find that works which focus on the suitability for metrology aspects report only on very low idler powers of only a few 10s of \si{mW}\,\cite{Balskus2016,Balskus:2016,Lee2014,Vainio2017}, which is several orders of magnitude lower than needed. In contrast, works that report on the highest average powers in this spectral region (\SI{7.8}{W} at \SI{3.57}{\micro\meter} for an OPO system\,\cite{Sudmeyer2004} and \SI{6.7}{W} at \SI{2.90}{\micro\meter} for a DFG based system\,\cite{Catanese2019}) do not sufficiently discuss metrology aspects, which is however critical to determine the suitability of the light sources for spectroscopy experiments. Additionally, several properties which are highly relevant for metrology applications, particularly a direct measurement of the idler PN or beam quality, have, to the best of our knowledge, not been discussed at all in literature for comparable systems.

In this work, we explore metrology aspects of an OPO system with very high output powers and discuss the suitability and perspectives of the OPO platform as a light source for power-hungry metrology applications in the mid-IR with a focus on cavity-enhanced Doppler-free saturation spectroscopy. To the best of our knowledge, we present the highest average power in the \SIrange[]{3}{5}{\micro\meter} region, reaching \SI{10.3}{W} for the idler output at \SI{3.1}{\micro\meter}. We analyze the PN of both signal and idler beam, and show that we were able to conserve the PN of the OPO pump. The onset of higher order modes (HOM) in the OPO cavity when scaling to high output powers is discussed and we differentiate between fundamental-mode and HOM operation of the OPO while presenting the respective properties as well as their suitability for metrology applications. The beam quality of both signal and idler is analyzed for both fundamental-mode and HOM operation. Additionally, we discuss spectral stability as well as the stability of the output power of the system. Our OPO features a broad wavelength tunable range from \SIrange[]{2700}{4700}{nm}, adjustable by translating the fanout-structured poling period of the used periodically-poled LiNbO$_3$ (PPLN) crystal. Our results enablethe realization of highly-stable, metrology-grade OFCs in the mid-IR regime and pave the way for precision spectroscopy techniques which have so far been limited by the available light source average powers and quality by reaching record high idler powers while at the same time ensuring low PN and excellent beam quality.

\section{Experimental Setup}
\subsection{OPO Setup and Pump}
\label{sec:setup}
\begin{figure}[t!]
    \centering
    \captionof{table}{Specifications of all optics used in the setup shown in Figure\,\ref{fig:setup}: RoC, radius of curvature; f, focal length; FS, frontside; BS, backside; R, reflectance for unpolarized light; T, transmittance for unpolarized light; Tp, transmittance for p-polarized light; Ts, transmittance for s-polarized light; AoI, angle of incidence. For all CMs, the FS faces towards the resonant signal beam, for all other optics the FS is the side where the pump and/or idler beam make the transition from air to substrate according to their path highlighted in Figure\,\ref{fig:setup}. The coatings specified as "UniNE" were custom designed and manufactured by V.J. Wittwer at Université de Neuchâtel. For all other coatings, the specifications of the respective suppliers were used.}
    \begin{tabular}{lllll}
        \hline
        Component & RoC/f & Coating & Substrate & Supplier/Part Number \\ 
        \hline
        CM1/CM2 & \SI{-209.1}{mm} & FS,BS: R $>$ 99.8\% \SIrange[]{1340}{1570}{nm} AoI = \SI{7}{\degree}, & CaF$_2$ & Substrate: \\ 
        ~ & ~ & FS,BS: T $>$ 98\% \SIrange[]{1018}{1064}{nm} AoI = \SI{7}{\degree}, & ~ & EKSMA Optics 112-5230E \\ 
        ~ & ~ & FS,BS: T $>$ 90\% \SIrange[]{2700}{5000}{nm} AoI = \SI{7}{\degree} & ~ & Coating: UniNE \\ 
        CM3  & plane & FS: R $>$ 99.5 \% \SIrange[]{1200}{1800}{nm} & UVFS & Laseroptik 12239\\ 
        CM4/OC  & plane & FS: R = 85\% ($\pm$ 3\%) \SIrange[]{1400}{1800}{nm}, & UVFS & Layertec 106828 \\ 
        ~ & ~ & BS: high T \SIrange[]{1400}{1800}{nm} & ~ & ~ \\ 
        PPLN & plane & FS,BS: T $>$ 99\% \SI{1040}{nm}, & 5\% Mg0 & HC Photonics \\ 
        ~ & ~ & FS,BS: T $>$ 99\% \SIrange[]{1310}{1650}{nm}, & LiNbO$_3$ & 1BKFL5P13250005100050Y002 \\ 
        ~ & ~ & FS,BS: T $>$ 90\% \SIrange[]{2800}{5050}{nm} & ~ & ~ \\ 
        DM  & plane & FS,BS: R $>$ 99.5\% \SIrange[]{1020}{1060}{nm} AoI = \SI{45}{\degree}, & CaF$_2$ & Substrate: EKSMA Optics \\ 
        ~ & ~ & BS: T $>$ 99.5\% \SIrange[]{3000}{4500}{nm} AoI = \SI{45}{\degree} & ~ & 530-5253; Coating: UniNE \\ 
        GW  & plane & FS,BS: Thorlabs C9 Coating, & Ge & Thorlabs WG91050-C9 \\ 
        ~ & ~ & T $\sim$ \SIrange[]{85}{95}{}\% \SIrange[]{1900}{6000}{nm} & ~ & ~ \\ 
        L1 & \SI{75}{mm} & FS,BS: Thorlabs YAG Coating, & UVFS & Thorlabs LA4725-YAG \\ 
        ~ & ~ & T $>$ 99.5\% \SIrange[]{1020}{1060}{nm} & ~ & ~ \\ 
        L2 & \SI{50}{mm} & same as L1 & UVFS & Thorlabs LA4148-YAG \\ 
        L3 & \SI{150}{mm} & FS,BS: Thorlabs E Coating, & CaF$_2$ & Thorlabs LA5012-E \\ 
        ~ & ~ & Total Average T $>$ 98.75\% \SIrange[]{2000}{5000}{nm} \\ 
        M1/M2 & plane & FS: R $>$ 99.9 \% \SIrange[]{1000}{1060}{nm} & UVFS & EKSMA Optics 042-1030HHR \\
        HWP1/HWP2 & plane & FS: T $>$ 99.5 \% \SI{1030}{nm} & UVFS & EKSMA Optics 464-4208 \\
        TFP & plane & Total Tp $>$ 98 \% \SI{1030}{nm} AoI = \SI{45}{\degree}, & UVFS & EKSMA Optics 420-1248i45HE \\
        ~ & ~ & Tp/Ts $>$ 1000:1 \\
        \hline
        ~ & ~ & ~ & ~ & ~ \\     
    \end{tabular}
    \label{tab:partlist}

      \centering
  \includegraphics[width=\textwidth]{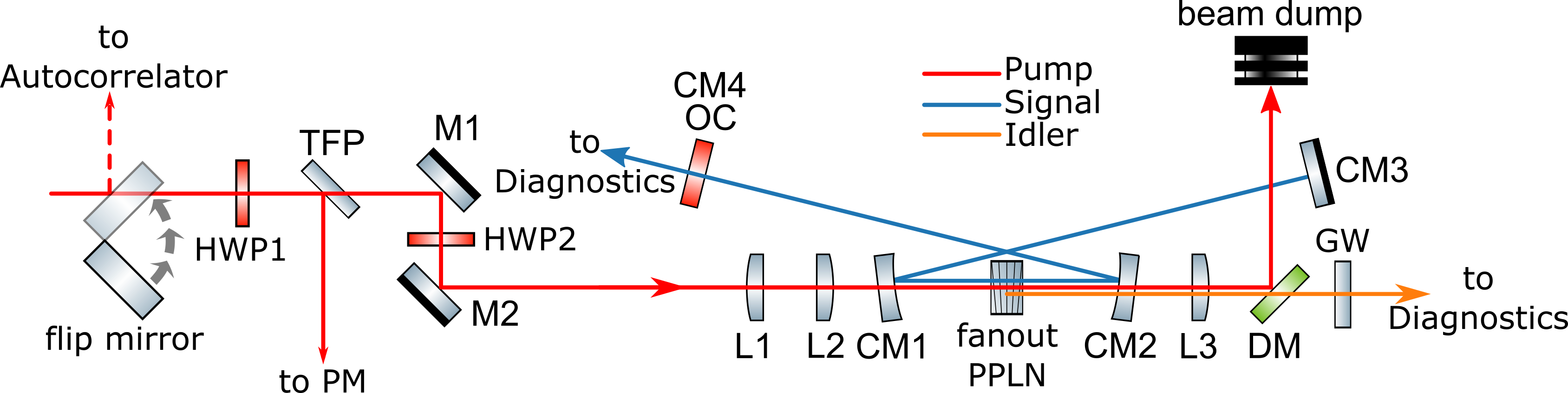}
\captionof{figure}{Scheme of the OPO setup: HWP1-2, half-wave plates; TFP, thin-film polarizer; M1-2, mirrors; PM, powermeter; L1-3, lenses; CM1-4, cavity mirrors; OC, 15\% output coupler; DM, dichroic mirror; GW, germanium window; PPLN, periodically-poled LiNbO$_3$ crystal.}
\label{fig:setup}
\end{figure}

A scheme of the OPO setup is shown in \textbf{Figure\,\ref{fig:setup}}. A more detailed description of the respective optical components is provided in \textbf{Table\,\ref{tab:partlist}}. The OPO is realized as a synchronously-pumped, signal singly-resonant, X-shaped standing-wave cavity. Cavity mirrors (CM) 1-2 are concave and broadband highly-reflective (HR) for the signal and highly transmissive (HT) for pump and idler. They are separated by \SI{260}{mm} with the nonlinear crystal placed centered inbetween. CM3 is a plane mirror which is HR for the signal and mounted on a translation stage to match the OPO cavity length to the pump repetition rate. CM4 serves as an output coupler (OC) for the signal. The geometrical distances $\mathrm{\overline{CM13}}$ and $\mathrm{\overline{CM24}}$ are $\sim$\,\SI{462}{mm}. The nonlinear crystal used is a \SI{5}{mm} long MgO(5\%):doped PPLN crystal, which is placed in an oven and operated at 70° \si{C}. The poling periods are organized in a fanout structure and can be tuned continuously from \SIrange[]{25.5}{32.5}{\micro\meter} by translating the PPLN normal to the incident beam path. The PPLN coating is optimized for pump and signal transmission, which leads to roughly 10\% idler losses at the exiting PPLN interface.

The pump is an inhouse-built mode-locked \SI{125.4}{MHz} low-noise Yb:fiber oscillator based on a nonlinear amplifying loop mirror (NALM)\,\cite{Mayer2020,Pecile2023} amplified by a chirped-pulse amplifier (CPA)\,\cite{Shumakova2022}, which delivers \SI{1042}{nm} centered, \SI{8}{nm} FWHM, $\sim$\SI{550}{fs} pulses at up to \SI{68}{W} of average power. The FWHM and the pulse duration are optimized to meet the spectral acceptance BW as well as the temporal walk-off length of the \SI{5}{mm} long PPLN\,\cite{Ramaiah-Badarla2014}. The pump beam is aligned to the OPO cavity via mirrors (M) 1-2 and mode-matched to the resonant signal beam via lenses (L) 1-2\,\cite{Baker1980}. The minimal signal waist radius centered inside the PPLN is estimated to be $\sim$\SI{80}{\micro\meter} using the ray transfer-matrix method for Gaussian beams\,\cite{Kogelnik1966}. This corresponds to a pump beam waist radius of $\sim$\SI{65}{\micro\meter}, the related M$^2$ measurement is shown in Section~\ref{sec:M2}. At the highest possible pump power of \SI{68}{W} the pump peak intensity inside the PPLN is calculated as \SI{14.86}{GW\per cm\squared}. This intensity is higher than reported laser-induced damage thresholds for lithium niobate using similar light sources\,\cite{Nikogosyan2005}, however, we did not observe any laser-induced damage in our crystal. The pump power can be adjusted by rotating half-wave plate (HWP) 1 and thereby changing the polarization-dependent transmission through the thin-film polarizer (TFP). HWP2 is used to create an s-polarized beam to align the pump polarization with the dipole moment of the PPLN. Both the residual pump and generated idler beams get collimated by L3 and subsequently separated by a dichroic mirror (DM). The residual pump is then dumped, the idler is transmitted through a germanium window (GW) to be cleaned from any leakage through the DM.

\subsection{Measurement Setup}
In this work we analyzed the optical spectra, average powers, PN, as well as the beam quality (M$^2$ and spatial beam profiles) of the pump, signal and idler beams. All spectra of the signal were recorded with a Yokogawa AQ6374 grating-based optical spectrum analyzer (OSA) with a resolution bandwidth (RBW) of \SI{2}{nm}. The spectra of the idler were measured with a Thorlabs OSA305 Fourier-transform OSA with a RBW of \SI{2.1}{nm}. 

The pump power was measured indirectly with a Thorlabs S425C-L powermeter (PM) by measuring the power of the beam reflected by the TFP and subtracting the measured value from the total power. The power of both signal and idler were directly measured with a Thorlabs S425C PM. 

The PN of pump and signal as well as all presented RF-traces were obtained with a \SI{5}{GHz} BW InGaAs photodetector (PD; Thorlabs DET08CL). The PN of the idler was measured with a \SI{1}{GHz} BW HgCdTe PD (Vigo Photonics UHSM-10.6). A Rohde \& Schwarz FSWP-8 was used for PN and RF trace analysis. All PN measurements presented in the script show $\mathscr{L}(f)\equiv S_\phi(f)/2=S_{SSB}(f)$ where $\mathscr{L}(f)$ is the standard measure of phase instability, $S_\phi(f)$ is the PN power spectral density (PSD), and $S_{SSB}(f)$ is the single sideband (SSB) PSD, following the current IEEE standard definitions\,\cite{IEEE2009}.
The PN traces presented in this work were measured at a higher harmonic $n$ of the repetition rate to increase the measurement sensitivity with respect to the shot-noise limit and then corrected to receive the PN of the repetition rate by subtracting the term $20 \log_{10} (n)$ from the whole PN trace, as similarly done in Modsching et al.\,\cite{Modsching2021}. The pump was measured at $n=16$, the signal at $n=19$ and the idler at $n=8$.

The spatial beam profiles were recorded and the M$^2$ measurement of the pump was performed with a Newport LBP2-HR-VIS2 laser beam profiler, for the signal and idler an Ophir Spiricon Pyrocam IV beam profiling system was used. The M$^2$ measurements were performed according to ISO 11146-1:2021.

\section{Results and Discussion}

\subsection{Fundamental Mode Operation and Spectral Tuning Range}
\label{sec:tunability}
\begin{figure}[htb!]
  \centering
  \includegraphics[width=\textwidth]{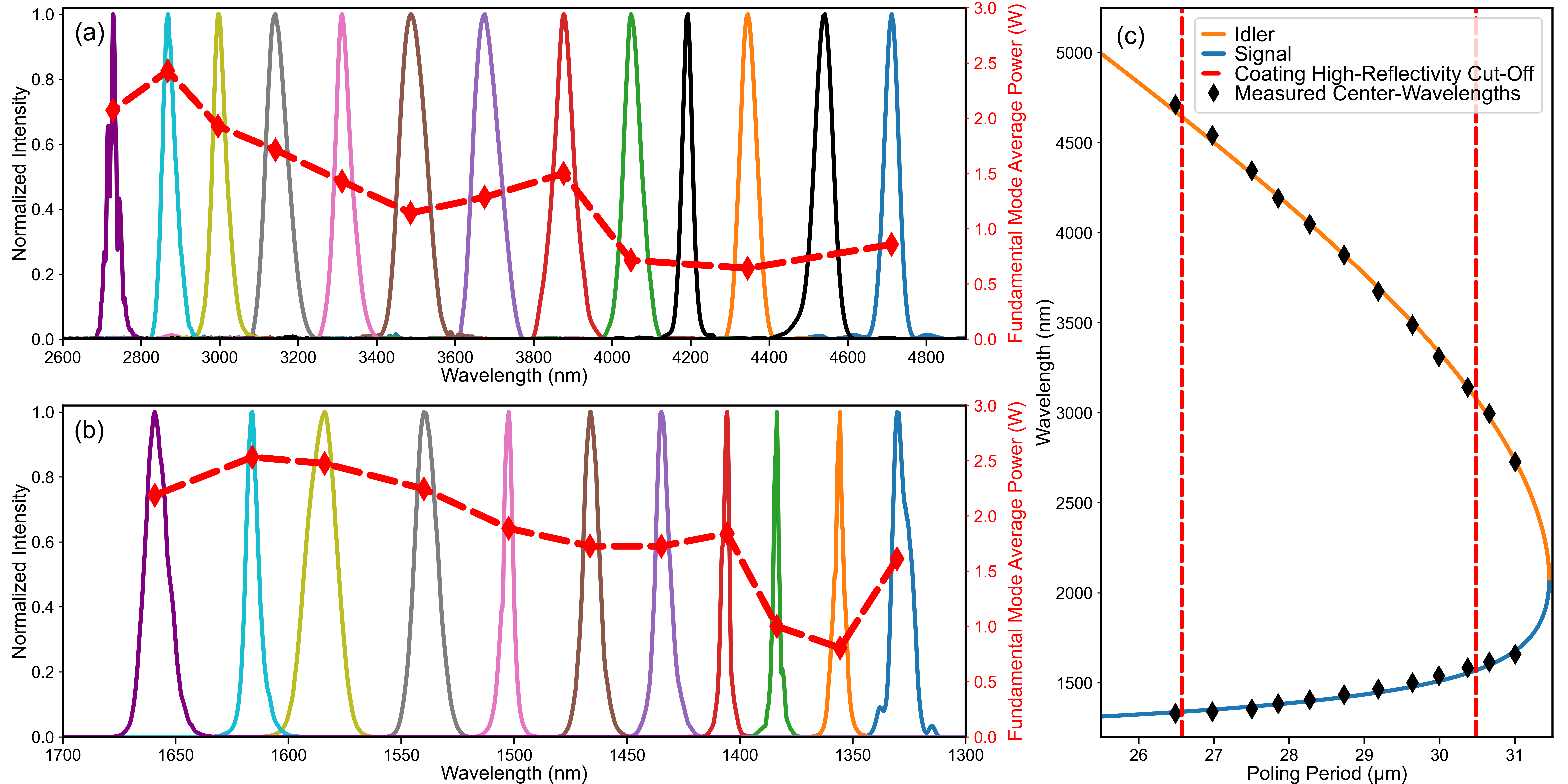}
\caption{Normalized spectra of the (a) idler and the (b) signal in fundamental-mode operation over the whole supported spectral tuning range. The spectra obtained at the same poling period are color-matched. Note, that the corresponding signal of the two idler spectra shown in black are not presented to omit overcrowding in the signal plot. The features for the idler at $\sim$\SI{2750}{nm} (purple) are attributed to CO$_2$ absorption in air. The red dotted lines in (a) and (b) show the highest achieved average power in  fundamental-mode operation prior to the excitation of HOM. The idler powers are corrected for the germanium window losses. (c) Calculated QPM curve for different poling periods of the used PPLN for the idler (orange) and signal (blue). The BW supported by the used coatings in the cavity (for details please refer to Table\,\ref{tab:partlist}; coating cut-offs shown in dashed red) can be exceeded slightly without significantly higher losses. The center wavelengths of the measured signal-idler-pairs (black diamonds) show good agreement with the theoretical QPM curve.}
\label{fig:spectra_SM}
\end{figure}

Measured spectra for the idler and signal in fundamental-mode operation over the whole spectral range are shown in \textbf{Figure\,\ref{fig:spectra_SM}}\,(a) and (b), respectively. The obtained average output powers for the signal and idler are depicted in the respective plots on the secondary y-axis. The calculated quasi-phase-matching (QPM) curve and determined center wavelengths of signal and idler are presented in Figure\,\ref{fig:spectra_SM}\,(c). \\Fundamental-mode operation was ensured by measuring a conservation of the pump PN for the signal and idler beams, as is discussed in greater detail in Section\,\ref{sec:PN}. We received bell-shaped spectra over the whole spectral range with a FWHM of \SIrange[]{45}{85}{nm} for the idler and \SIrange[]{5}{15}{nm} for the signal. The OPO threshold was in the range of \SIrange[]{300}{500}{mW} for idler wavelengths below \SI{4000}{nm} and increased gradually for wavelengths larger than \SI{4000}{nm}, reaching $\sim$\SI{1800}{mW} at \SI{4700}{nm}. The photon-to-photon conversion efficiency (PCE) was calculated to be $\sim$\SI{40}{}\% for idler wavelengths of \SIrange[]{2700}{3900}{nm}. Beyond \SI{4000}{nm}, \SIrange{35}{10}{}\% could be achieved, decreasing with higher wavelengths. These values are similar to PCEs reported for other signal singly-resonant OPO systems operating in the same spectral region, where \SIrange{35}{50}{}\% are achieved for wavelengths smaller \SI{4000}{nm} before gradually dropping down to $\sim$\SI{10}{}\% for higher idler wavelengths\,\cite{Nagele2020,Adler2009}. The decreased performance for higher idler wavelengths is expected, as the OPO process experiences less gain when moving away from degeneracy\,\cite{Cerullo2003,Butterworth1993,Heckl2016}. Additionally, PPLN crystals show increased absorption above \SI{4000}{nm}\,\cite{Nikogosyan2005}.

The highest average power in fundamental-mode operation achieved for the idler was \SI{2.43}{W} at a wavelength of \SI{2870}{nm}. The lowest average power was obtained as \SI{0.64}{W} at \SI{4345}{nm}. To best of our knowledge, those are the highest average powers reported over the whole tuning range of our system (\SIrange[]{2700}{4700}{nm}), compared to other wavelength tunable parametric sources operating in the same wavelength regime\,\cite{Nagele2020,Adler2009}. In general, the obtained idler powers in fundamental-mode operation become smaller with higher wavelengths. This is explainable with two effects: First, the average powers for higher wavelengths are smaller, as the energy per photon decreases with increasing wavelength. Second, when moving away from degeneracy, the OPO process experiences less gain, as already discussed above.

For the signal, the highest fundamental-mode average power was measured as \SI{2.53}{W} at \SI{1615}{nm}, the lowest as \SI{0.81}{W} at \SI{1355}{nm}. Here, we see the same trend as for the idler: When moving away from degeneracy, we achieve lower average powers. However, here the difference is smaller compared to the idler: As we move towards shorter wavelengths, the photons carry more energy and higher average powers are achieved even when lower gains are present.

\subsection{Higher-Order-Mode Behavior}
\label{sec:MM}
\begin{figure}[htb!]
  \centering
  \includegraphics[width=\textwidth]{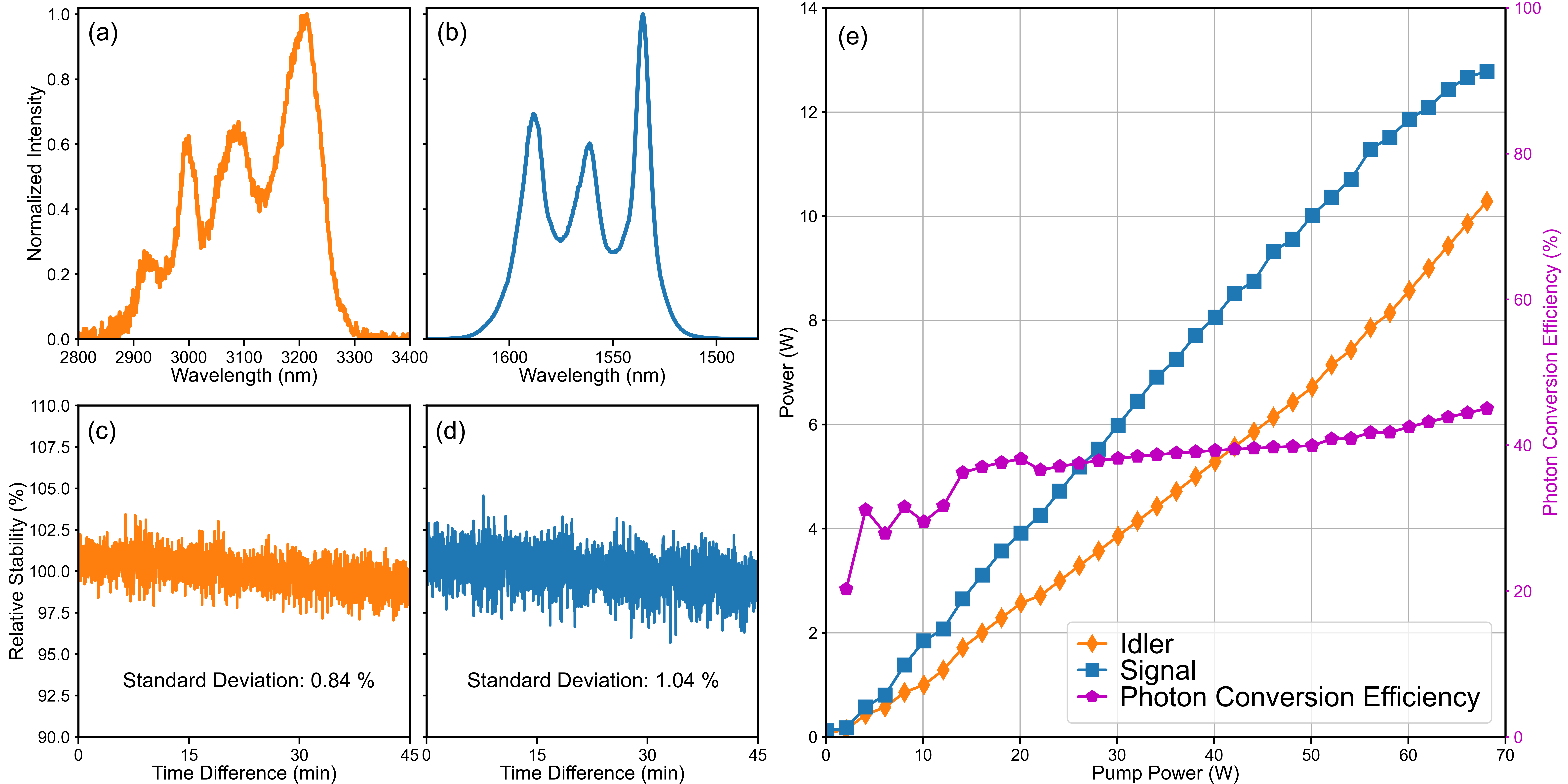}
\caption{Normalized spectrum of the (a) idler and the (b) signal at full pump power in stable HOM operation. \SI{45}{min} average power stability measurement of the idler (c) and the signal (d). The single measurement values are normalized to the mean value. The sampling interval was \SI{0.5}{\second}.(e) Pump power slope of signal and idler. The idler power values are corrected for the germanium window transmission losses. The OPO was realigned at \SI{50}{W} pump power to compensate for thermal effects, which lead to a steeper slope efficiency for the highest pump powers. The slight decrease in slope efficiency for high signal powers is caused by operating the used PM outside of it's \SI{10}{W} maximal power rating, which leads to a decreased response.}
\label{fig:MM_stab_ps}
\end{figure}

When the pump power is increased beyond the fundamental-mode limit, HOMs are excited in the OPO cavity. In the following, we discuss the HOM behavior exemplarily for an idler wavelength of \SI{3100}{nm}. The corresponding power slope is shown in \textbf{Figure\,\ref{fig:MM_stab_ps}}\,(e). We achieve \SI{10.3}{W} idler average power at a pump power of \SI{68}{W}, which, to the best of our knowledge, is the highest average power in the \SIrange{3}{5}{\micro\meter} region so far reported\,\cite{Catanese2019,Sudmeyer2004}. The PCE amounts to 45\% at maximal pump power, which is slightly higher than the results received for fundamental-mode operation. We also achieve a signal average power of \SI{12.8}{W} at \SI{1550}{nm} for maximal pump power, which is among the highest powers reported in that specific wavelength region\,\cite{Sudmeyer2004}. The output power stability of the system was recorded and the standard deviation (SD) calculated over \SI{45}{\minute}, showing a stability of 0.84\% for the idler and 1.04\% for the signal (see Figure\,\ref{fig:MM_stab_ps}(c), (d), respectively). The used sampling interval was \SI{0.5}{\second}. These results are surprisingly good, as we only see slightly higher instabilities compared to values reported for low-power systems well below \SI{100}{mW}, which achieved idler stabilities of 0.62\% in free-running operation\,\cite{Nagele2020} and 0.30\% with active cavity length stabilization\,\cite{Vainio2017}, even though we operate at more than 100 times higher power. The signal and idler powers show a linear slow-term drift over the full measurement time which is attributed to the walk-off of the pump and signal pulses due to cavity length fluctuations. In the presented configuration both the seed oscillator as well as the OPO were free-running, with the OPO being operated in an open lab environment. A further improvement in stability can be expected by applying, e.g., active cavity length stabilization schemes or passive thermal shielding. 

The recorded optical spectra at full pump power of the idler and the signal are shown in Figure\,\ref{fig:MM_stab_ps}\,(a) and (b), respectively. They are significantly more broadband as the fundamental-mode spectra shown in Figure\,\ref{fig:spectra_SM}\,(a) and (b) (the idler expands roughly over \SI{400}{nm}, the signal almost over \SI{100}{nm}), featuring several distinct peaks. For stable operation in the HOM regime a realignment of the OPO cavity right after reaching the desired pump power and again after a thermalization period (usually around \SI{30}{\minute}) was critical to achieve stable operation. Without realignment, mode-beating can be observed and both the spectra of signal and idler as well as their respective output powers become instable.

\subsection{Phase Noise Analysis}
\label{sec:PN}
\begin{figure}[htb!]
  \centering
  \includegraphics[width=\textwidth]{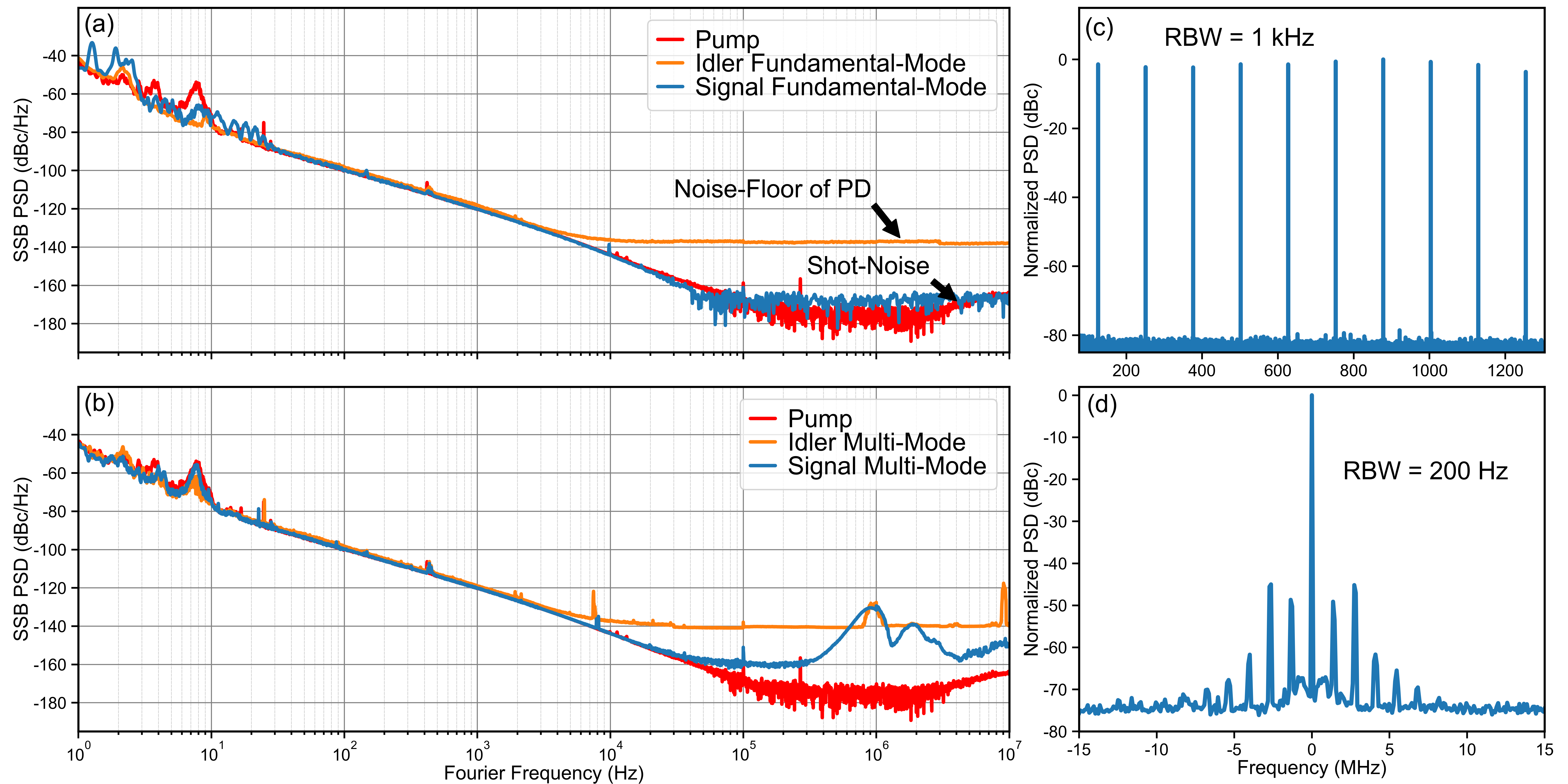}
\caption{Phase noise analysis of the OPO in fundamental-mode (a) and HOM onset (b) operation. We see, that the PN of the pump is conserved in case of fundamental-mode operation for both signal and idler. The pump and signal measurements are limited by photon shot-noise at high Fourier frequencies, while the idler measurement is limited by the intrinsic noise of the used PD. In the case of HOM operation, we see additional features appearing for both signal and idler. (c) RF spectrum of the signal in the case of HOM onset. Even though distortions are already clearly visible in the PN (the trace corresponds to the data presented in (b)), no modulations can be seen in the RF trace. (d) Zoom into the fundamental harmonic of the repetition rate for the signal in the case of stable HOM operation at a pump power significantly above the HOM threshold.}
\label{fig:PN_RF}
\end{figure}

The PN of the repetition rate of pump, signal and idler in fundamental-mode operation is shown in \textbf{Figure\,\ref{fig:PN_RF}}\,(a). We see, that the pump PN is conserved over the full range of Fourier frequencies. The here exemplarily presented PN measurements have been performed with the idler set to \SI{3300}{nm} and the signal to \SI{1580}{nm}. Both pump and signal PN are limited by photon shot-noise for Fourier frequencies larger \SI{40}{kHz}. In the case of the idler, PN conservation can only be guaranteed up to $\sim$\SI{10}{kHz}, as we are here limited by the intrinsic noise floor of the used PD before reaching shot-noise. However, we can observe a very good agreement of all curves for all Fourier frequencies within their respective measurement limitations.

The changes in PN for HOM onset are shown in Figure\,\ref{fig:PN_RF}\,(b). The pump PN can not be conserved in this case. The first changes in PN which usually appear in the case of HOM onset are very narrow peaks appearing as additional features in the measured traces. These peaks represent beating between the fundamental mode and HOM due to their different longitudinal mode number\,\cite{Lee2002}. In the here presented traces they can be seen at Fourier frequencies of $\sim$\SI{8}{kHz} and $\sim$\SI{9}{MHz}. Additionally, broad features in the Fourier frequency range of a few \SI{100}{kHz} to a few \si{MHz} are typical, as shown here for both the signal and idler traces.

Changes in PN could already be observed, when all other diagnostic means used throughout this study would still indicate fundamental-mode operation, i.e., no degradation in beam quality, or change in the optical spectrum as well as the RF spectrum of the PDs could be noticed. For example, in Figure\,\ref{fig:PN_RF}\,(c) we show the RF spectrum of the signal that corresponds to the PN trace presented in (b), resembling a clean Dirac-comb up to a SNR of $>$\SI{80}{dB}, where no modulations or any other perturbations can yet be observed.
Only when the pump power is further increased beyond the HOM onset threshold, the HOM behavior of the OPO can also be directly observed in the RF traces, as these modulations become clearly visible, which we show exemplarily for the fundamental harmonic of the signal in Figure\,\ref{fig:PN_RF}\,(d).

Summarizing, we could show that signal and idler PN are preserved in the case of fundamental-mode operation of the OPO cavity. A HOM onset can already be detected by PN measurements before any instability in the optical spectrum or modulations in the RF spectrum as well as a beam quality degradation can be observed. PN measurements therefore are identified as the most sensitive means to assess fundamental-mode stability of an OPO cavity and detect HOM onset early on. Additionally, we could show that signal and idler show the same PN performance. This enables for a prediction of the temporal properties of the idler based solely on a characterization of the near-infrared signal beam, where detectors are cheap and available off-the-shelf, while mid-IR PDs are still scarcely available and expensive.

\subsection{Beam Quality Measurements}
\label{sec:M2}
\begin{figure}[htb!]
  \centering
  \includegraphics[width=\textwidth]{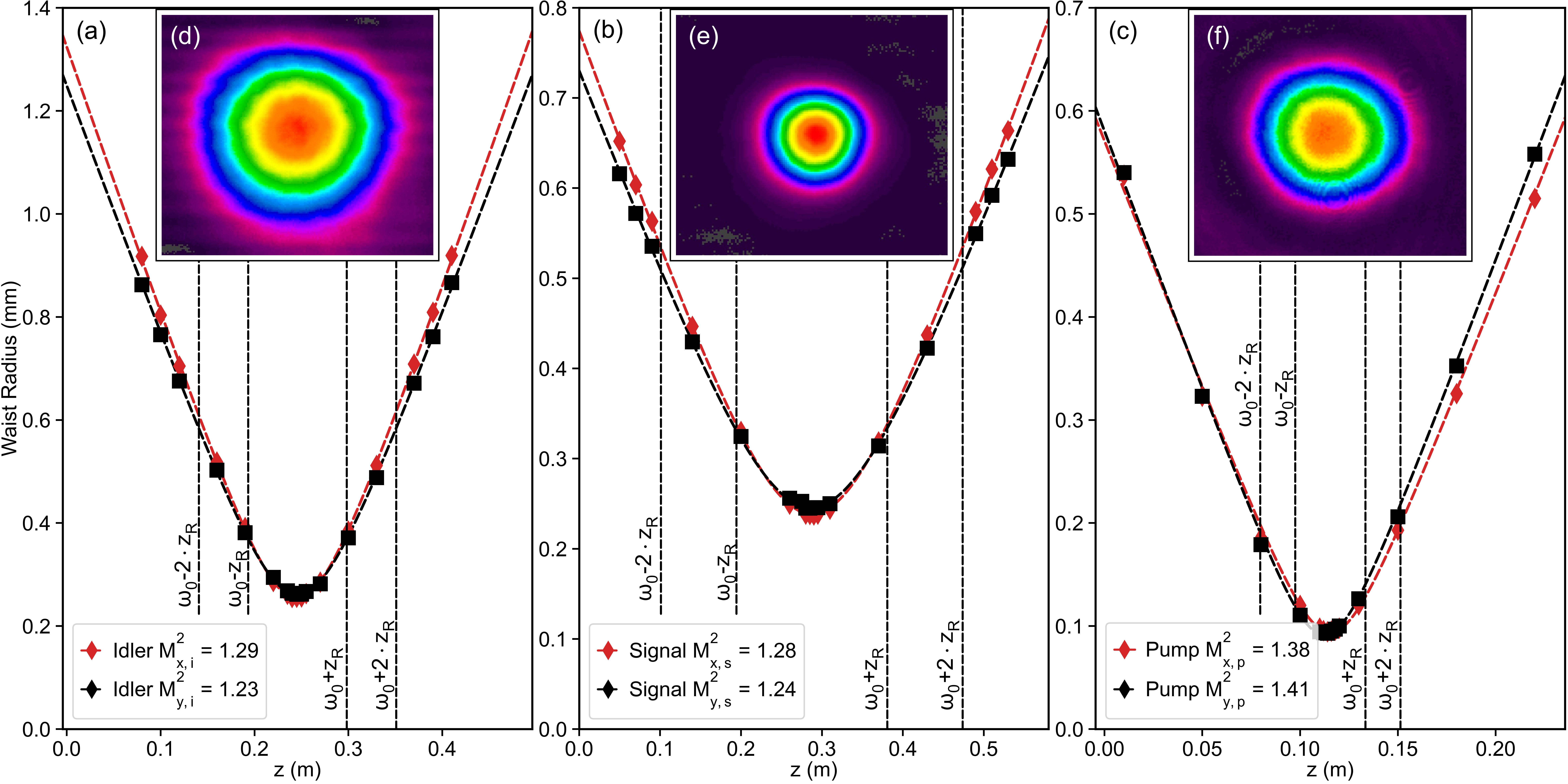}
\caption{Fundamental-mode M$^2$ measurement for (a) idler, (b) signal and (c) pump. The pump measurement was performed between CM1 and CM2 (with removed PPLN) and therefore corresponds to the mode-matching configuration used in the here presented work. The indices \textit{x} and \textit{y} refer to the horizontal and vertical planes, respectively, \textit{i}, \textit{s} and \textit{p} refer to the idler, signal and pump beams. The distances of one, respectively two Rayleigh lengths ($\mathrm{z_R}$) from the minimal waist positions ($\mathrm{\omega_0}$) are shown as black dashed lines. Collimated beam profiles of fundamental-mode idler (d), signal (e) and pump (f). Please note, that the "fringes" seen in inset (d) are caused by the chopper of the used beam profiler (Pyrocam IV) and are a measurement artefact.}
\label{fig:M2_BP}
\end{figure}

M$^2$ measurements for idler and signal in fundamental-mode operation are shown in \textbf{Figure\,\ref{fig:M2_BP}}\,(a) and (b), respectively. The M$^2$ measurement of the pump beam is shown in Figure\,\ref{fig:M2_BP}\,(c). The pump measurement was performed between CM1 and CM2 (with removed PPLN) and therefore resembles the OPO cavity mode-matching configuration used in our setup. The corresponding collimated beam profiles are shown in Figure\,\ref{fig:M2_BP}\,(d), (e) and (f). The here discussed measurements were performed at an idler wavelength of \SI{3300}{nm} and a signal wavelength of \SI{1500}{nm}.

For the idler we received values of $\mathrm{M_{x,\,i}^2=1.29}$ in the horizontal and $\mathrm{M_{y,\,i}^2=1.23}$ in the vertical plane. For the signal we obtained $\mathrm{M_{x,\,s}^2=1.28}$ and $\mathrm{M_{y,\,s}^2=1.24}$, which shows very good agreement with the values retrieved for the idler. In both cases, the beams have slightly lower $\mathrm{M^2}$ values in the vertical plane. This can be explained by the horizontal tilt of CM1 and CM2 relative to the resonant signal beam wavefront and the thereby introduced beam ellipticity, which is also predicted by calculations using the ray transfer-matrix method for the sagittal and tangential planes separately\,\cite{Kogelnik1966}. 

For the pump beam we received values of $\mathrm{M_{x,\,p}^2=1.38}$ and $\mathrm{M_{y,\,p}^2=1.41}$. We see, that both signal and idler beam show a better beam quality, i.e., a lower M$^2$ value than the incident pump beam. Additionally, the slight astigmatism of the pump beam (see Figure\,\ref{fig:M2_BP}\,(c)) can not be observed for both signal and idler. We attribute this to the nature of the parametric process itself: Areas with high intensity contribute stronger to nonlinear conversion and therefore only parts of the pump beam close to the center contribute, leading to "cleaned" signal and idler beams.

\section{Conclusion and Outlook}
A high-power low-PN OPO system with a focus on its suitability for power-hungry metrology applications was investigated. We could show, that parametric systems can reach very high idler output powers, achieving the record-high average power of \SI{10.3}{W} at \SI{3100}{nm} with an excellent power stability of 0.84\% (SD). We discussed that very high output powers in OPOs lead to HOM onset, differentiated between fundamental-mode and HOM behavior of the OPO, and highlighted the different properties of both operational regimes. Fundamental-mode operation, which is critical for cavity-enhanced spectroscopy applications, was achieved for up to \SI{2.43}{W} of average power at \SI{2870}{nm} reporting, to the best of our knowledge, the highest average powers over the whole tuning range of our system (\SIrange[]{2700}{4700}{nm}), compared to other wavelength tunable parametric sources operating in the same wavelength regime. Additionally, we have shown that we were able to conserve the low PN of the pump beam and achieve excellent beam quality ($\mathrm{M^2}<1.3$) while maintaining narrow spectral widths of the idler beam. To study the suitability of OPOs as platform for high-power mid-IR metrology applications, we investigated several parameters, which have not been reported on before in literature. Examples are a direct measurement of the PN of the mid-IR idler beam as well as it's beam quality aspects and interconnections between the signal and idler beam properties. In our here presented work we could also identify PN measurements as the most sensitive diagnostic means to assess fundamental-mode stability and thereby as critical to identify HOM onset early on.

For future systems, we suspect that metrology-grade mid-IR OFCs based on OPOs (i.e. systems that can provide a high simultaneous stability of average power and spectral shape, as well as a low M$^2$ value and low phase noise) with even higher output power could be achieved by increasing the cavity OC rate and thereby shifting the HOM onset to higher powers. This approach has been successfully implemented by Südmeyer et al., who used very high OC rates of $\sim$\,\SI{96}{}\% in combination with a very long, high-gain nonlinear crystal\,\cite{Sudmeyer2004}. However, for very high idler average powers in the \SI{10}{W} regime or beyond, a combination of an OPO system with a subsequent OPA amplification stage seems to be a more promising approach. This has already been shown for low-power mid-IR systems\,\cite{Nagele2020}. This way the necessity of very high gain can be avoided, which is known to compromise beam quality\,\cite{Sudmeyer2004,Arisholm2004} and low M$^2$ values can be maintained even for very high powers.

Summarizing, we presented a highly-stable, low phase noise, metrology-grade OPO system with record-high powers in the \SIrange[]{3}{5}{\micro\meter} wavelength regime, thereby facilitating advanced metrology and spectroscopy applications previously constrained by power limitations and insufficient light source stability\,\cite{Shumakova2024}.

% Acknowledgements
\medskip
\textbf{Acknowledgements} \par
This research was funded in whole or in part by the Austrian Science Fund (FWF) [DOI: 10.55776/P36040 \\and 10.55776/F1004]. For open access purposes, the author has applied a CC BY public copyright license to any author accepted manuscript version arising from this submission. The financial support by the Austrian Federal Ministry for Digital and Economic Affairs, the National Foundation for Research, Technology and Development and the Christian Doppler Research Association is gratefully acknowledged. The LTF would like to acknowledge its funding: Schweizerischer Nationalfonds zur Förderung der Wissenschaftlichen Forschung (SNSF) (206021\_198176 and 206021\_170772).

\medskip
\textbf{Conflict of Interest} \par
Vito F. Pecile, Michael Leskowschek and Oliver H. Heckl receive funding from Thorlabs Inc.

\medskip
\textbf{Data Availability Statement} \par
Data underlying the results presented in this paper are not publicly available at this time but may be obtained from the authors upon reasonable request.

% References
\medskip

% Use the following code if you wish to generate your bibliography with BibTeX;
% replace the string "MSP-template" below with the name(s) of
% the BibTeX data base(s) you want to use.
% The resulting bibliography-output (the content of the .bbl file)
% must be pasted back into this file before submission.
% Please also include your BibTeX data base file(s) in your submission
% so that we can re-run BibTeX if necessary.
%
\bibliographystyle{MSP}
\bibliography{bibliography}

\end{document}